\documentstyle[aps,epsfig,harvard,fancyheadings]{revtex}

\newcommand\be{\begin{equation}}
\newcommand\ee{\end{equation}}

\begin{document}


\title{Topological and energetic factors: what determines the
structural details of the transition state ensemble and ``on--route" intermediates
for protein folding? An investigation for small globular proteins.}

\author{Cecilia Clementi, Hugh Nymeyer, Jos\'e Nelson Onuchic}

\address{Department of Physics, University of California at San Diego, \\
La Jolla, California 92093-0319, USA}

\date{\today }
\maketitle

\vspace{1cm}


\begin{abstract}
Recent experimental results suggest that the native fold, or topology,
plays a primary role in determining the structure of the transition
state ensemble, at least for small fast folding proteins.  To
investigate the extent of the topological control of the folding
process, we study the folding of simplified models of five small globular
proteins constructed using a G\=o--like potential in order to retain
the information about the native structures but drastically reduce the
energetic frustration and energetic heterogeneity among
residue--residue native interactions. By comparing the structure of the
transition state ensemble experimentally determined by $\Phi$--values
and of the intermediates with the ones obtained using our models, we show that
these energetically unfrustrated models can reproduce the global
experimentally known features of the transition state ensembles and ``on--route"
intermediates, at least for the analyzed proteins.  
This result clearly indicates that, as long as the protein
sequence is sufficiently minimally frustrated, topology plays a
central role in determining the folding mechanism.

\vspace{1cm}


\end{abstract}

\twocolumn
\section{Introduction}

Our understanding of the protein folding problem has been thoroughly
changed by the new view that has emerged in the last decade.  This new
view, based on the energy landscape theory and funnel concept 
\cite{Leopold92,Bryngelson95,Socci96,Onuchic97,Dill97,Nymeyer98,Klimov96,Shakhnovich96:FD,Shea98},
describes folding as the progressive evolution of an ensemble of partially folded
structures through which the protein moves on its way to the native structure.
The existence of a deep energy funnel in
natural proteins and the relatively simple connectivity between most
conformational states which are structurally close makes
this description possible even when only a few simple reaction
coordinates that measure similarity to the native structure are used.
The folding mechanism is controlled by both the shape of this free
energy landscape and the roughness on it, which arises from the
conflicts among interactions that stabilize the folded state and therefore
can create non--native conformational traps
\cite{Bryngelson87,Bryngelson89,GoldsteinRA92}.

The energetic roughness, however, is not the only limiting factor in
determining a sequence's foldability.  Even if the energetic roughness
could be completely removed, the folding landscape would not be
completely smooth. Theoretical
\cite{WolynesPG96,Nelson97,Nelson98,Onuchic96,Socci97a,Betancourt95,Sheinerman98,Micheletti99,ScheragaHA92:prosci}
and experimental \cite{Baker98,Serrano98} 
advances indicate that the
final structure of the protein also plays a major role in determining
a protein's foldability.  Some particular folding motifs may be
intrinsically more designable than others. To address this difference
in foldability which is not dependent on energetic frustration, we
have introduced the concept of ``topological frustration''
\cite{Nymeyer99:PNAS,Onuchic99,Shea99}.

Let us imagine an ideal situation for which the order of native
contact formation during folding is not biased.  In this ``ideal''
situation, there are an enormously large number of equivalent folding
pathways, and an analysis of the transition state ensemble would show
that for this ensemble nearly all parts of the protein have a similar
probability of participation.  The structure in the transition
ensemble has been estimated by analogy with minimalist lattice models
made to reproduce the global landscape features of small, fast
folding proteins: similar Levinthal entropies, stabilities and 
energetic roughness as gauged by the glass transition temperatures.
These models show a transition state ensemble about half way through
the unfolded and folded states \cite{Onuchic95}. In this ideal case,
all the contacts in this transition ensemble would exist with
the same probability.

Although the average amount of native formation in the transition
ensemble is about 50\%, the lattice simulations show that, even
when the sequence is designed to have substantially reduced energetic
frustration, there are variations in the amount of nativeness of
specific contacts in the transition state ensemble 
\cite{Onuchic96,Onuchic99,Nymeyer99:PNAS}. Real
proteins display similar heterogeneity in contact formation.  
In systems with no energetic frustration and equal native interactions, 
these variations 
in the transition state ensemble are created solely by the folding motif 
and polymeric constraints that make certain contacts more geometrically 
accessible and stable than others.  This variation in frequency 
that some contacts are made in the transition state ensemble
generally reduces the entropy of
the transition state and, when determined by the native motif, is a
gauge of the amount of ``topological frustration'' in the system.
Although this type of frustration can be modified by some design
tricks \cite{plotkinnew}, it cannot be completely eliminated: it
reflects an intrinsic difficulty in folding to a particularly chosen
shape.  Minimalist models have shown how this heterogeneity
leads to a transition ensemble that is a collection of diffuse
nuclei which have various levels of native contact participation
\cite{Onuchic96}.  The minimalist models calibrated to real proteins
show similar overall levels of contact heterogeneity as in real
proteins \cite{Onuchic96}.  This picture of a transition state composed of
several diffuse nuclei has been confirmed by other lattice and
off--lattice studies \cite{KlimovD98,Pande&Rokhsar98}.  In addition to
selecting sequences which have low levels of energetic frustration,
evolution appears to have selected for a particular set of folding motifs
which have reduced levels of ``topological frustration'', discarding
other structures to which it is too difficult to fold
\cite{Betancourt95,Wolynes96,Nelson98,Micheletti99,Debe99}.

Guided by theoretical folding studies on lattice, off--lattice, and
all--atom simulations (see for instance
\citeasnoun{Onuchic95},
\citeasnoun{Onuchic96},
\citeasnoun{Boczko95},
\citeasnoun{Onuchic99},
\citeasnoun{Nymeyer99:PNAS},
\citeasnoun{Shea99})
as well as recent experimental evidence
\cite{Baker98,Serrano98,DobsonCM99:NSB,SerranoL99:NSB,Baker99:NSB},
we suggest that real
proteins, and especially small, fast folding (sub--millisecond),
two--state like proteins, have sequences with a sufficiently reduced
level of energetic frustration that the experimentally observed
``structural polarization'' of the transition state ensemble ({\it
viz.\/} the variation in the amount of local native structure) is
primarily determined by the topological constraints.  That is, in
well designed sequences, the variations are more determined by the
type of native fold than by differences in sequence which leave
the native fold relatively unchanged and the energetic frustration
small.  

The amount of native local structure in the transition state can be
experimentally measured by using single and double point mutants as
probes in the $\Phi$ value technique \cite{Fersht94:COSB}.  If the
topology is a dominant source of heterogeneity in
transition state structure, then the majority of evolved sequences
which fold to the same motif would exhibit similar local structure in
the transition state ensemble.  We provide evidence in this paper that
not only is this the case, that much of the transition state ensemble
is determined by the final folded form, but, also for larger proteins
that are not two--state folders, some ``on--route'' intermediates are
determined by topological effects as well.  Thus it appears that the
dominance of topology in folding extends even into some larger, slower
folding proteins with intermediates. This fact is consistent with some
recent observations by Plaxco and collaborators that reveal a
substantial correlation between the average sequence separation
between contacting residues in the native structure and the folding
rates for single domain proteins \cite{Plaxco98:JMB,Chan98:Nat}.

To ascertain the extent of topological control of the folding
behavior, we create several simplified energetic models of small, 
globular proteins using potentials created to minimize energetic
frustration. We show that these energetically unfrustrated models
reproduce nearly all the known global features of the transition states
of the real proteins on whose native structures they are based,
including the structure of folding intermediates.  We directly compare
the structure of the transition state ensemble experimentally 
determined by $\Phi$ value measurements with the numerically 
determined one.  The simulated transition state ensemble is inferred 
from structures sampled in
equilibrium around the free energy barrier between the folded and unfolded states.
This free energy is computed as a function of a
single reaction coordinate that measures the fraction of formed native
contacts. The validity of this method has been demonstrated in references
\cite{Onuchic99,Nymeyer99:PNAS}. 

The organization of the paper is as follows: in section \ref{sec2} we
present in some detail the physical concepts underlying this work in
the light of recent experimental results.  In section \ref{sec3} we
present results for a sample of five small, globular proteins,
and compared these results against the available experimental data.
The off--lattice model used in our study is presented in the Appendix.
In order to investigate the relevance of the topology, we chose a
model which reproduces the topological features of a given real
protein and eliminates most of the energetic frustration and
variations in the strength of native residue--residue contacts.  The
predicted transition state for these proteins are in good agreement
with experimental evidences, supporting our hypothesis of the major
role played by topology.

\section{Checking the Folding Mechanism by Analyzing the Transition State 
Ensemble}               
\label{sec2}

How do we know what the folding transition state ensemble looks like?
Experimental analysis of folding transition state ensembles has been
largely performed using the $\Phi$--value analysis technique
introduced by Fersht and co--workers \cite{Fersht94:COSB}.  $\Phi$
values measure the effects that a mutation at a
given position along the chain has on the folding rate and stability:
\be
\label{phi_exp}
\Phi \equiv \frac{-R T \ln (k_{mut}/k_{wt})}{\Delta \Delta G^{0}}
\ee
where $k_{mut}$ and $k_{wt}$ are the mutant and wild--type folding rate
respectively, $R$ is the ideal gas constant, $T$ is the absolute
temperature, and $\Delta \Delta G^{0}$ is the difference in the total
stability between the mutant and wild--type proteins in $kcal/mol$.

Because the folding event of small fast folding proteins is well
described as a diffusive process over a barrier determined by the free
energy profile, the folding rate can be written as a Kramer's--like
equation
\cite{Socci96}
\begin{equation}
k = k_0 \exp [ -\Delta G^{\ddag}/ R T ]
\end{equation}
where $k_0$ is a factor depending on the barrier shape and the
configurational diffusion coefficient of the system.  If $k_0$ is
insensitive to small sequence changes, what appears to be true for
reasonably unfrustrated sequences 
\cite{Onuchic96,Socci96,Nymeyer99:PNAS,Shea99,Onuchic99,Baker97b,Munoz99:PNAS} 
the $\Phi$ value is then seen to be a ratio of free energy changes of the 
folding barrier to stability:
\begin{equation}
\Phi = \frac{\Delta \Delta  G^{\ddag}}{\Delta \Delta G^{0}}
\label{therm_phi}
\end{equation}
where $\Delta \Delta G^{\ddag}$ is given by
\begin{equation}
\Delta \Delta  G^{\ddag} = \Delta G^{\ddag}_{mut} - \Delta
G^{\ddag}_{wt} = -R T \ln \frac{k_{mut}}{k_{wt}}.
\end{equation}
When this relationship is valid and the mutation can be considered a
small perturbation, the $\Phi$ value is a convenient measure of the
fraction of native structure which is formed in the transition state
ensemble around the site of the mutation.  A $\Phi$ value close to 1
means that the free energy change between the mutant and the wild type
is almost the same in the transition state and native state,
indicating that native contacts involving the mutated residue are
already formed at the transition state.  Inversely, a $\Phi$ close to
$0$ means that the free energy change is the same in the transition
state and unfolded states, so the local environment of the residue is
probably unfolded--like.  A detailed analysis of the mutation is
needed to determine exactly what contacts are disrupted under
mutation.  Ideally, mutations are made which eliminate small
hydrophobic side--groups.  Studies using $\Phi$ values with multiple
same--site mutations generally support the accuracy of $\Phi$ value as
a structural measurement of the transition ensemble
\cite{Matouschek95}, although sizable changes in the transition state
structure have been induced in at least one protein through a single
point mutation \cite{Oas97}.  In interpreting $\Phi$ values, it is
also important to remember that they only measure the relative change
in structure, not the absolute amount of structure.  This leads to the
possibility that some mutants with low $\Phi$ values may have nearly
native local environments in the transition state, a possibility seen
clearly in the experimental studies of {\it Procarboxypeptidase
A2}~\cite{Serrano98:JMB}.

The validity of $\Phi$ values as structural measurements clearly
supports the Kramer's--like description of the folding rate and the
fact that the $\Phi$ can be properly understood as a ratio of the free
energy change of the transition ensemble over the change of the native
ensemble (equation \ref{therm_phi}).  This latter equation is very
convenient as a starting point for computing $\Phi$ values.  In
several recent simulation papers for lattice and off--lattice protein
models, we have investigated this issue at length
\cite{Nymeyer99:PNAS,Onuchic99,Shea99}. All these studies concluded
that as long as the systems present a weak or moderate level of
energetic frustration (such as the G\=o--like models in this work),
$\Phi$ values determined from changes in the free energy barrier,determined
using a single simple reaction coordinate, yield quantitatively
correct $\Phi$ values. Therefore, all the calculations performed in this
work were done utilizing eq.~\ref{therm_phi} --- no actual
kinetics was performed but only the appropriate sampling of the protein
configurational space (see Appendix and
refs.~\citeasnoun{Socci95},\citeasnoun{Boczko95},
\citeasnoun{Onuchic96},\citeasnoun{Nymeyer99:PNAS}, for example,
for details). Technically, as long as the folding barriers are of a few
$k_B T$ or more and the displacement of the barrier position along this
reaction coordinate under mutation is sufficiently small,  the
$\Phi$ values can be computed using free energy perturbation:
\begin{equation}
\Phi = \frac{\Delta \Delta G^{TS} - \Delta \Delta G^{U}}{\Delta \Delta
G^{F} - \Delta \Delta G^{U}} = 
\frac{\ln \langle e^{\Delta E / R T} \rangle_{TS} - 
\ln \langle e^{\Delta E / R T} \rangle_{U}} 
{\ln \langle e^{\Delta E / R T} \rangle_{F} - 
\ln \langle e^{\Delta E / R T} \rangle_{U}}.
\label{simple_phi}
\end{equation}
We use equation \ref{simple_phi} to compute $\Phi$ values for our
protein models using fixed transition, unfolded, and folded regions
identified by the free energy profile viewed using a single order
parameter: $Q$, the fraction of native contacts formed in a given
conformation.

What experimental evidence exists as to the role of topology in
determining the average structure in the folding transition state
ensemble?  The clearest evidence to date of the role of topology comes
from comparisons of the transition state structure of two homologues
of the {\it SH3} domain ({\it src SH3} and {\it $\alpha$--spectrin SH3}). 
These two homologues have only weak identity ($\approx 30$\% identity 
with gaps), but $\Phi$ values at corresponding sequence positions are
highly correlated \cite{Baker98,Serrano98}, supporting the
degeneracy in the folding behavior for these two sequences.
Furthermore, one of these sequences has a strained $\Phi$--$\Psi$
conformation in the high $\Phi$ region of the distal turn.  The fact
that this strain does not detectably lower the $\Phi$ values in the
local neighborhood \cite{Serrano98}, suggests that the sequence
details and local stability are less important for determining how
structured a region is in the transition state ensemble than its
location in the final folded conformation.  Other evidence indicates
that these results may be more generally applicable than simply for {\it SH3}
or $\beta$--sheet proteins.  Sequence conservation has been shown not
to correlate with $\Phi$ values \cite{Baker98:PNAS}, indicating that
in general sequence changes at a given position in a protein weakly
affect the $\Phi$ value at that position.

Results for some small fast folding proteins (such as {\it CI2} and
the $\lambda$--{\it repressor}) suggest that the transition state is an
expanded version of the native state, with a certain degree of
additional inhomogeneity over the structure \cite{Itzhaki95,Oas97}
(similar to the theoretical predications for small $\alpha$--helical
proteins \cite{Onuchic95,Boczko95}), while results for other proteins
(as the $\beta$--sheet {\it SH3} domain) show apparently larger
structural heterogeneity in the transition state
\cite{Sheinerman98,Sheinerman98a}. This difference in the degree of
``structural polarization'' that is emerging between small
$\alpha$--helix and $\beta$--sheet proteins suggests that the folding
mechanism of a given protein is fundamentally tied to the type of
secondary structural elements and their native arrangement.  Current
studies using $\Phi$ value technique have been made of {\it src SH3}
\cite{Baker98}, $\alpha$--{\it spectrin SH3} \cite{Serrano98}, {\it
CI2} \cite{Itzhaki95}, {\it Barnase} \cite{Fersht92}, {\it Barstar}
\cite{Fersht99}, $\lambda$--{\it repressor} \cite{Oas97}, {\it CheY}
\cite{LopezHernandez96}, {\it protein L} \cite{Baker98:JMB}, {\it
Procarboxypeptidase A2}~\cite{Serrano98:JMB}, {\it RNase H}~\cite{Raschke99}
and the tetrameric
protein domain from tumor suppressor {\it p53}~\cite{Mateu99:NSB}.

In this paper, we analyze five proteins ({\it SH3}, {\it CI2},
{\it Barnase}, {\it RNase H} and {\it CheY}) that have been 
extensively studied experimentally and for which, therefore, details 
of their transition state ensemble are quite well known. 
We generate sequences (and potentials) for simulating these
different globular proteins.  These sequences have the native backbone
folds of real experimentally studied globular proteins but
sequence and potential interactions designed to drastically reduce the
energetic frustration and heterogeneity in residue--residue
interactions.  By comparing the transition state structures of these
unfrustrated models with the experimental studies of their real protein
cousins, we quantify the effects of the native topology.  If
topology completely determines how folding occurs, then the model
and real proteins should have identical folding behavior and $\Phi$
values.  If energetic frustration and heterogeneity are critical for
determining the folding mechanism, then the variations in $\Phi$
values with position should bear far reduced similarity to those 
in the real proteins on which the computer homologues are based.

Two of the five studied proteins
are simple two--state like fast folding proteins ({\it SH3} and {\it
CI2}), while  the other three ({\it Barnase}, {\it RNase H} and 
{\it CheY}) are known to fold through
the formation of an intermediate state.  We show not only that our
simple models can reproduce most of the $\Phi$ value
structure, but also that models for {\it Barnase}, {\it RNase H} and
{\it CheY} correctly
reproduce the folding intermediates of these proteins, suggesting
that many of the ``on--route'' intermediates are also largely
determined by the type of native fold.

We represent the five globular proteins using a simplified C$_\alpha$
model with a G\=o--like~\cite{Ueda75} Hamiltonian as detailed in the
Appendix.  This potential is in its details unlike that of real
proteins, which have residue--residue interactions with many
components (Coulomb interactions, hydrogen bonding, solvent mediated
interactions, etc., etc.).
The crucial features of this potential are its low level of energetic
frustration, that characterizes good folders and a native conformation
equal to the real protein. The ability of this
model to reproduce features of the real transition state ensemble and
real folding intermediates is a strong indication that the retention
of the topology is enough to determine the global features of their 
folding mechanism.  Using these models, we
simulate the dynamics of a protein starting from its native structure,
for several temperatures.  To monitor the thermodynamics of the
system, we group the configurations obtained during a simulation as a
function of the reaction coordinate, $Q$, defined as the fraction of
the native contacts formed in a conformation ($Q=0$ at the fully
unfolded state and $Q=1$ at the folded state).  The choice of $Q$ as
order parameter for the folding is motivated by the fact that in a
funnel--like energy landscape, a well designed sequence has the
energy of its conformations reasonably correlated to degree of 
nativeness, and the parameter
$Q$ is a good measure of the degree of similarity with the native
structure.  Our G\=o--like potential is minimally frustrated for the
chosen native structure, and the prediction of transition state
ensemble structures and folding rates for these G\=o--like systems has
been shown to be quite accurate \cite{Socci96,Shea99,Nymeyer99:PNAS}.  From
the free energy profile as a function of $Q$, it is easy to locate the
unfolded, folded and transition state ensembles, as it is shown in next
section.
Since these models consider totally unfrustrated sequences,
they may not reproduce the precise energetics of the real proteins, 
such as the value of the barrier heights and the stability of the
intermediates, nonetheless they are able to determine the general 
structure of these ensembles. 

In order to compare the folding process simulated using our model to
the actual process for a given protein (as obtained from experimental
$\Phi$--values analysis), we need to choose a ``mutation'' protocol to
compute $\Phi$ values.  Experimentally, the ideal mutation is
typically one that removes a small hydrophobic side--group such as a
methyl group that makes well--defined and identifiable residue--residue
contacts in the native state.  The $\Phi$ value is then sensitive to
this known contact.  Our computational mutation is the removal of a
single native bond, so our computer $\Phi$ values are sensitive to the
fractional formation of this bond $Q_{ij}$ between residues $i$ and
$j$.  We make these mutations because, as in most real mutations, they
are sensitive to the formation of specific contacts, rather than being
averages over interactions with many parts of the native structure.
They mostly resemble the interaction $\Phi_{int}$ value made by making
double cycle mutants \cite{Fersht92}.  $\Phi$ values are computed from
equation \ref{simple_phi}.  
In an ideal, perfectly smooth funnel--like energy landscape, all the 
$\Phi$ values should be equal; in an energetically unfrustrated
situation, $\Phi$ values 
variations are due to the structure of the native conformation.

\section{Determining the Transition State Ensemble of Small Globular Proteins}
\label{sec3}

We have discussed the idea of ``topological frustration'' and its role
in determining the structural heterogeneity of the transition state
ensemble.  We explore its role directly by creating protein models
which drastically reduce the energetic frustration and energetic
heterogeneity among residue--residue native interactions leaving the
topology as the primary source of the residual frustration.  Results
obtained with these models, constructed using a C$_\alpha$ level of
resolution with a G\=o--like potential designed to fold to the
native trace of chosen proteins, are then compared against the
experimental data of those proteins. Five proteins with different
folding motifs and different amounts of transition state heterogeneity
(variation in $\Phi$ values) and/or intermediates have been investigated.

We first analyze {\it Chymotrypsin Inhibitor II} ({\it CI2}), a mixed
$\alpha$--$\beta$ protein with a broad distribution of $\Phi$ values
(nearly uniform from 0 to 1).  Then we present an analysis for the {\it
src SH3} domain, a largely $\beta$--sheet protein with a more polarized
transition state structure (a substantial number of large $\Phi$
values).  We then apply the same technique to {\it Barnase}, {\it
Ribonuclease H} ({\it RNase H}) and {\it CheY}, three other
mixed $\alpha$--$\beta$ proteins which fold via a folding intermediate.
Although these proteins are not two--state folding proteins, we
demonstrate that topology is also the dominant determinant of their
folding behavior.  We show that the topology plays a major role not
only in the transition state ensemble, but it is also largely
responsible for the existence and general structure of the folding
intermediate.  This result may be quite common for ``on--route"
folding intermediates and could provide a computational method for
distinguishing between ``on--pathway" and ``off--pathway" structures which are
inferred from experiments. 
To check the applicability of this method, the same approach presented in 
this paper has been extended elsewhere~\cite{Clementi:PNAS_submitted} to a pair 
of larger proteins ({\it Dihydrofolate Reductase} and {\it Interleukin}--1$\beta$). 
Even for these very large proteins we found that the overall structure of the
transition state and intermediate ensembles experimentally observed
can be obtained utilizing similar simplified models.

\subsection{Analysis of two--state folders: CI2 and SH3}

\subsubsection{CI2}

The {\it Chymotrypsin Inhibitor 2} ({\it CI2}) protein is a 64 residue
protein, consisting of six $\beta$--sheets packed against an
$\alpha$--helix to form a hydrophobic core.
Experimental studies
\cite{Fersht91b,JacksonSE91ii,Jackson93a,JacksonSE93} have established
that {\it CI2} folding and unfolding can be modeled by simple
two--state kinetics.  The structure of the transition state for this
protein has been extensively characterized by protein engineering
\cite{Itzhaki95,Otzen95,Fersht91b}, by free energy functional
approaches \cite{Shoemaker99a,Shoemaker99b}, by a geometrical
variational principle \cite{Micheletti99}, and by all--atom molecular
dynamics simulations \cite{Daggett96,Daggett99,Karplus97}.  These studies have
shown the transition state has roughly half of the native interactions
formed in the transition state ensemble and a broad distribution of
$\Phi$ values in agreement with the general predictions of the energy
landscape theory used with a law of corresponding states for small
proteins \cite{Onuchic95,Onuchic96}. The broad distribution of $\Phi$
values suggests that most hydrophobic contacts are represented at a
level of about 50\% in the transition state ensemble.

We constructed a G\=o--like C$_\alpha$ model of {\it CI2} as described
in the Appendix.  Several fixed temperature simulations were made
and combined using the WHAM algorithm \cite{Swendsen93} to generate a
specific heat versus temperature profile and a plot of the potential
of mean force as a function of the folding order parameter $Q$
(see figure \ref{fig1_ci2}).  From the free energy profile,
we identified the dominant barrier, and used the thermal ensemble of
states at its location to generate $\Phi$ values from equation
\ref{simple_phi}.  The ranges of values of $Q$ used 
to determine each of these ensembles are shaded in figure
\ref{fig1_ci2}.  The mutations have been implemented by the removal of
single attractive interactions (they are replaced with the same short
ranged repulsive interactions used between residues without native
interactions).  The values computed via this method are shown in
figure \ref{fig2_ci2}.  Also shown in this figure is the fractional
formation of individual native contacts in the transition state.  The small
difference between these two figures is primarily due to the fact that
in the $\Phi$ calculations the native contact formation at the folded and
unfolded states are also taken into account.
Because of the higher concentration of contacts between residues
near--by in sequence and the local conformational preferences, the 
unfolded state shows a high level of local structure. The
inaccurate representation of local contacts in the unfolded state
makes the short range $\Phi$ values less reliable as transition
structure estimates than long range $\Phi$ values.  

From the calculations, we detect three significant
regions of large $\Phi$ values: the $\alpha$--helix, the
mini--core defined by strands 3 and 4 and their connecting 
loop, and between the C--terminus of strand 4 and the N--terminus of 
strand 5. These regions
generally have $\Phi$ values in excess of $0.6$.  Slightly smaller
values of about $0.5$ exist for the short range contacts between the
N--terminal of strand 3 and the C--terminal of the $\alpha$--helix and
for contacts between strand 3 and strand 4.  All other regions lack
a consistent set of large $\Phi$ values.  Despite the large number of
native contacts between strands 1 and 2 and the $\alpha$--helix and
between strands 5 and 6 and the $\alpha$--helix, only low $\Phi$
values are observed in this region (nearly all below $0.2$ in value).
A comparison between these data and the exhaustive analysis of Fersht and
colleagues \cite{Otzen95} shows excellent overall agreement. They have
found that ``$\beta$--strands 1, 5 and 6 ... are not structured in the
transition state....''.  Strand 2 also shows a highly reduced amount of
structure.  Furthermore, ``the central residues of $\beta$--strands 3
and 4 interact with the $\alpha$--helix to form the major hydrophobic
core of CI2.''  The hydrophobic mini--core in this region (defined as
the cluster formed by side--chains of residues 32, 38, and 50) is
detected by single mutant and double mutant $\Phi$--values
\cite{Itzhaki95} to be at least 30\% formed in transition ensemble.
Similarly, they found the $\alpha$--helix, particularly the N--capping
region, to be highly ordered.

In summary, we see a quite good overall agreement except for a
discrepancy in the short range interactions in the loop region between
strands 4 and 5.  This protein shows generally 
higher $\Phi$ values between interactions
which are more local in sequence and lower $\Phi$ values between
interactions which are distant in sequence.  The results are thus
consistent with the picture of the transition state as a collection of
non--specific and somewhat diffuse nuclei \cite{Onuchic95}.  
This overall low level of frustration
suggests a low level of ``topological frustration'' in this model as
well and a particularly designable motif.

\subsubsection{src SH3 domain}
\label{sh3_sec}

{\it Src SH3} is the 57 residue fragment of {\it Tyrosine--Protein
Kinase} that stretches from T84 to S140.  It has five $\beta$ strands
(and a short 3--10 helix) in an anti--parallel arrangement, forming a
partial $\beta$ sandwich. Experimental measurements have shown that 
the {\it SH3} domain
folds using a rapid, apparently two--state mechanism.  A $\Phi$ value
analysis \cite{Baker98} reveals that the distal loop hairpin and
diverging turn regions are both highly structured and docked together
at the transition state; the hydrophobic interactions between the base
of the hairpin and the strand following the diverging turn are
partially formed, while other regions of {\it src SH3} appear only
weakly ordered in the transition state ensemble.  The overall
representation of the transition state structure of {\it src
SH3} ---having the distal loop and diverging turn largely formed and
other regions weakly formed--- agrees with studies of $\alpha$--{\it
spectrin SH3},~\cite{Serrano98} which has a similar backbone
structure but a dissimilar sequence (approx 30\% identity with gaps).
This observed similarity along with evidence of a strained backbone
conformation in the distal loop of the $\alpha$--{\it spectrin SH3}
\cite{Serrano98} supports the concept of ``topological'' dominance in
folding \cite{Baker98}.

Fig. \ref{fig1_sh3} shows the folding behavior as obtained
from our dynamics simulations of the G\=o--like analogous of the {\it src SH3}.
The free energy barrier defining the transition state location is 
evident in the figure.
As before, we have computed $\Phi$ values from equation
\ref{simple_phi} by mutating (removing) every native residue--residue
attractive contact. The results of this calculation are shown in
figure \ref{fig2_sh3}. In addition to $\Phi$ values, the contact
formation probability at the transition state ensemble have been calculated.
Our previous caveats concerning $\Phi$ values for local interactions
still apply.  We observe the highest collection of off--diagonal 
(long range) $\Phi$ values
is in the diverging turn ---distal loop interaction exactly as seen
from the experimental $\Phi$ value measurements.  We see very low
values in the RT loop region, in accord with the two mutants in this
loop.  We also see medium to high values between the two $\beta$
strands which are connected by the distal loop.  The transition state
structure of the {\it SH3} presents a substantially larger degree of
structural polarization than {\it CI2}, where the $\Phi$ values are
much more uniform.  This suggests that {\it SH3} has a backbone
conformation which is intrinsically more difficult to fold, i.e.,
there is a greater level of ``topological frustration'' in this
structure.  Nevertheless the transition state composition is well
reproduced for both the two proteins.

\subsection{Analysis of three proteins which fold throughout the 
formation of an intermediate state: Barnase, RNase H and CheY}

%
{\it Barnase}, {\it RNase H} and {\it CheY} are three small
$\alpha$--$\beta$ proteins (although larger than the previous two
proteins):
{\it Barnase} is a 110 residue protein, composed by three 
$\alpha$--helices (located in the first 45
residues) followed by five $\beta$--strands; {\it RNase H} consists of 155 
residues which arrange themselves in five $\alpha$--helices and 
five $\beta$--strands; {\it CheY} is a 129 residues, classic $\alpha\/\beta$ 
parallel fold in which five $\beta$--strands are surrounded by five 
$\alpha$--helices.  
%
%
Experimental results show that these three proteins do not
fold by following a simple two--state kinetics directly from the
unfolded state to the native structure, but fold through the
formation of a metastable intermediate which interconverts
into the native state.  This brings up an interesting question: is
topology alone able to determine the presence of an intermediate in
the folding process?  In Figs. \ref{fig1_brn}, \ref{fig1_rnase} and 
\ref{fig1_chey} we show evidence for the
first time that such intermediates can be created solely from a
G\=o--like minimalist model which preserves the native topology.
The presence of this intermediate during these protein's folding
events is a requirement of the native protein motifs.
The free energy changes upon mutations of a wild--type three--state protein
are experimentally measured both for the intermediate and the
transition state, to define two different sets of $\Phi$--values for
the protein: 
\be
\begin{array}{c}
\Phi_{I} = \frac{\Delta \Delta G_{I} - \Delta \Delta G_{U}}
                {\Delta \Delta G_{F} - \Delta \Delta G_{U}}
\\
\Phi_{TS} = \frac{\Delta \Delta G_{TS} - \Delta \Delta G_{U}}
                 {\Delta \Delta G_{F} - \Delta \Delta G_{U}}
\end{array}
\ee
where $\Phi_{I}$ provides information about the structural composition of the 
intermediate state (I), and $\Phi_{TS}$ of the transition state (TS).
Following we discuss in some details the results for the three proteins.
Since, as for the first two proteins, the $\Phi$--values and the native contact
probabilities provide somewhat similar information, for simplicity, we show only the
results obtained for the native contact probabilities (for safety we have checked the
$\Phi$--values and determined that similar information is recovered).


\subsubsection{Barnase}

The analysis of experimentally obtained $\Phi$ values \cite{Fersht92} for
the {\it Barnase} shows that some relevant regions of the structure are fully 
unfolded in the intermediate while other regions are fully folded.

Fig. \ref{cm_stat.brn.fig} shows the intermediate and the transition
state structure obtained from the G\=o--like model.
The intermediate shows substantial structural heterogeneity:
there are very high probability values for interactions within the
$\beta$--sheet region and its included loops, and very low 
values for interactions within the $\alpha$--helices and their loops
and between the $\alpha$--helical and $\beta$--sheet regions.  Some
local short range helical interactions are formed.  The transition
state ensemble structure shows the same structure as the intermediate
with the addition of strong interactions within helices 2 and 3;
between helix 2, helix 3, the first $\beta$--strand, and the
intervening loops; and between the second $\beta$--strand and the
second helix.  

Comparing these simulation results with extensive mutagenesis studies of
reference \cite{Fersht92}, we observe a good qualitative agreement.  The
$\beta$--sheet region is highly structured in the intermediate as it is
the core region 3 (consisting of the packing of loop 3, that joins 
strands 1 and 2, and of loop 5, that joins strands 4 and 5, with
the other side of the $\beta$--sheet). In agreement with experiments, 
the earliest formed part of the protein appears to be the 
$\beta$--sheet region.  Also the core region 2 (formed by the
hydrophobic residues from helix 2, helix 3, the first strand, 
and the first two loops) is found to be only weakly formed in 
the intermediate and the transition state.  

There are two minor discrepancies between the {\it Barnase} model 
and the experimental data.  First, we slightly overestimate the formation 
of core region 2 in the transition state ensemble. Second, we 
underestimate the amount of structure in core
region 1 (formed by the packing of the first helix against a side 
of the $\beta$--sheet) in both intermediate and transition ensemble.  
In particular, we under--represent the interaction between helix 1 and the
$\beta$--sheet region. The experimentally observed early packing of
helix 1 against the rest of the structure is not reproduced by our
model. Clearly there are some important energetic factors which have
been neglected by the simple model. These may be inferred from
the {\it Barnase} crystal structure. For example, one can see that
helix 1 is largely solvent exposed, with interactions between it and
the remainder of the protein formed by only five of the eleven helix
residues.  83 \% of the interactions reside on the hydrophobic
residues PHE7, ALA11, LEU14 and GLN15, and the 17 \% of the
interactions are formed by the charged residues ASP8 and ASP12, while
the solvent exposed part of the helix is composed of polar residues.
Large stabilizing interactions other than tertiary
(most hydrophobic) interactions are neglected in the model, being 
probably responsible
for the failure in predicting the formation of the structural parts
involving helix 1. In this structural detail, it appears that the 
topological factors are not the leading determinant of the folding behavior.

\subsubsection{Ribonuclease H}

Kinetic studies of the wild--type {\it RNase H} have shown that an
intermediate state is populated in the folding process, and the structure
of this intermediate has been extensively investigated 
by circular dichroism, fluorescence and hydrogen exchange methods 
\cite{DaboraMarqusee94,Yamasaki95,DaboraMarqusee96,Chamberlain96,RaschkeMarqusee97} 
and by protein engineering \cite{Raschke99}. 
Fig. \ref{fig1_rnase} shows that, consistently with the experimental evidences,
we find an intermediate state in the folding process of the {\it RNase H} 
model. 
Experimental results indicate that the most stable region of the
protein intermediate involves the $\alpha$--helix 1, the strand 4, 
the $\alpha$--helix 4 and the $\alpha$--helix 2. Hydrogen exchange experiments 
have shown that the $\alpha$--helix 1 is the region of the protein most 
protected from exchange, suggesting that most of the interactions 
involving the $\alpha$--helix 1 are already significantly formed at the 
intermediate state of the folding process.  
The helix 4 and the $\beta$--strand 4 are the next most protected regions,
while the $\alpha$--helix 5 has low to moderate level of protection.
After the completion of the this intermediate structure, the 
rate--limiting transition state involves the ordering of the
$\beta$--sheet and the $\alpha$--helix 5. The packing of helix 5 across the
sheet is found to be the latest folding event.

The results of the model for {\it RNase H} show a good agreement with 
the experimental evidences. As shown in Fig. \ref{fig_rnase_cmap}, we find that the
formation of contacts involving the helix 1 is the earliest event in the
folding process.  Contacts arising from the $\alpha$--helix 4 and 
the $\beta$--strand 4 are then formed at the intermediate state and 
consolidated at the transition state.
In agreement with the experimental results, we find that, at the transition state, 
interactions between the $\alpha$--helix 1, the strand 4 
and the rest of the protein are mostly formed; the $\alpha$--helix
4 is also well structured and interactions between the helix 4 and the other 
parts of the protein are partly formed.
Interactions among the strands are almost all formed, but the sheet is
not yet docked to the helix 5.  
     
\subsubsection{CheY}

Utilizing protein engineering 
\cite{LopezHernandez96,LopezJMB97}, the transition state of {\it CheY} has been 
characterized and it can be described as a combination of two subdomains: the 
first half of the
protein (subdomain 1), comprising the $\alpha$--helices 1 and 2 and 
the $\beta$--strands 1--3, is substantially folded whereas the second half
(subdomain 2) is completely disorganized. The helix 1 seems
to play the role of a nucleation site around which subdomain 1 begins
to form. Moreover, an intermediate has been detected at the early stage 
of the folding process where all the five $\alpha$--helices are rather 
structured. The last two helices, however, are very unstructured in the 
later occurring transition state. From this result it has been suggested 
that a misfolded species is visited at the beginning of the folding process. 

Our simple model detects two possible intermediates for this protein, one of them
is an ``on--route" intermediate that is short--living and occurs just before the
transition state ensemble ($Q$ around 0.6 in Fig. \ref{fig1_chey}).
Surprisingly, the unfrustrated model is also able to detected 
a ``misfolded'' trap in the folding of {\it CheY}. Since non--native 
interactions are not allowed in the model, this trap is a 
long--living partially folded state created by the topological constrains.
There is no direct connection between this trap state and the fully folded state. 
The structure of this trap is shown in Fig.
\ref{fig_chey_cmap} and it agrees with the experimental observation of all helices well
structured. Differently 
from the previously discussed proteins, the model of {\it CheY} seems to have a 
tendency to first form a ``wrong'' part of the protein and, when this 
happens, a partial unfolding must occur before the folding can be 
completed. 

Finally, analyzing the transition state structure, we find 
a good agreement with the experimental data.
As shown in Fig. \ref{fig_chey_cmap}, the
first part of the protein (subdomain 1) is almost fully folded at
the transition state ensemble, while subdomain 2 is completely unfolded.

\section{Conclusions}

Recent theoretical studies and experimental results suggest that the
folding mechanism for small fast folding proteins is strongly
determined by the native state topology. The amount of energetic
frustration, arising from the residual conflict among the amino--acid
interactions, appears largely reduced for these proteins so that
topological constraints are important factors in governing the folding
process. Towards exploring this topological influence in real
proteins, we analyzed the folding process of the G\=o--like analogous of
five real proteins.  Since we have used G\=o--like potentials, the
energetic frustration is effectively removed from the system, while
the native fold topology is taken into account. It is important to
highlight that the results from such studies exhibit the overall
topological features of the folding mechanism, although we do not
expect the precise energetic values for barrier heights and
intermediate state stabilities. For example, real proteins are not
necessarily totally unfrustrated and they have only to minimize
energetic frustration to a sufficiently reduced level in order to be
good folders. Also, as long as energetic frustration is small enough,
creating some heterogeneity at the native interactions may help to
reduce topological frustration \cite{plotkinnew}, and that will
energetically favor some contacts over others.

The effective use of a small number of global order parameters as
reaction coordinates, in interpreting real data or studying more
detailed protein folding model, depends critically on the degree of
frustration present in real proteins \cite{Nymeyer99:PNAS}.  Since our
results show that general structural features of the transition state
ensemble in real proteins, at least for this class of fast folding
proteins, is reproducible by using a substantially unfrustrated
potential, several different global order parameters should work to
explain the folding mechanism. For this reason, it should not be a
surprise the fact that, utilizing energy landscape ideas and the
funnel concept, some very simple models with approximate order
parameters determined by single or few sequence approximation
\cite{Alm99:PNAS,Munoz99:PNAS,Galzitskaya99:PNAS} have been successful
in predicting qualitative features of the transition state ensemble.

Again, we have compared in details the structure of the transition state
ensemble of the five proteins resulting from our simulations with
experimental data. The agreement between our results and the
experimental data supports the idea that energetic frustration is
indeed sufficiently reduced and the protein folding mechanism, at
least for small globular proteins, is strongly dependent on
topological effects.  The structure of the transition state ensemble
of the {\it CI2} presents a broad distribution of $\Phi$ values
---i.e. a reduced degree of structural polarization--- in agreement with
predictions based on the energy landscape theory (see
\citeasnoun{Onuchic95}, \citeasnoun{Onuchic96}). On the other hand,
the structure of the {\it SH3} transition state ensemble shows a
higher degree of polarization. Nevertheless, by using our simplified
G\=o--like model, we have reproduced the transition state composition for
both proteins, demonstrating that topology is largely responsible for
the observed experimental differences.  The last three proteins we have
analyzed, ({\it Barnase}, {\it RNase H} and {\it CheY}) are known to 
fold through a three--state kinetics, involving the formation of an 
intermediate structure. Our G\=o--like model of these proteins also fold 
with a three--state kinetics with intermediates that are analogous 
to the ones detected experimentally. This fact suggests that topology 
is also a dominant factor in determining the ``on--route" intermediates.

\section{acknowledgments}

We thank Viara Grantcharova and David Baker for informations about the
{\it SH3} structure. We also thank Vladimir Sobolev for the CSU software.
We are indebted to Angel Garc\'{\i}a, Peter Wolynes, Steve Plotkin, 
Jorge Chahine, Joan Shea, Margaret Cheung, Charlie Brooks, Amos Maritan 
and Jayanth Banavar for helpful discussions. 
One of us (C.C.) expresses her gratitude to Giovanni Fossati for his 
suggestions and for carefully reading the manuscript, and to the 
Center for Astrophysics \& Space Sciences of UCSD
for the usage of graphics facilities and computer time. This work has 
been supported by the NSF (Grant \#96--03839), by the La Jolla
Interfaces in Science program (sponsored by the Burroughs Wellcome
Fund) and by the Molecular Biophysics training grant program 
(NIH T32 GN08326).

\newpage
\begin{appendix}
\section*{Model and Method}
\label{appendix}

In order to investigate how the native state topology affects the folding
of a given protein we follow the dynamics of the protein by using a G\=o--like 
Hamiltonian \cite{Ueda75} to describe the energy of the protein in a given
configuration. A G\=o--like Hamiltonian takes into account only native 
interactions, and each of these interactions enters in the energy balance
with the same weight.  It means that the system gains energy as much as any 
amino acid pair involved in a native contact is close to its native 
configuration, no matter how strong the actual interaction is in the real 
protein. 
Residues in a given protein are represented as 
single beads centered in their C--$\alpha$ positions. Adjacent beads 
are strung together into a polymer chain by mean of bond and angle 
interactions, while the geometry of the native state is encoded 
in the dihedral angle potential and a non--local potential. The energy of
a configuration $\Gamma$ of a protein having the configuration $\Gamma_0$ 
as its native state is thus given by the expression:
\be
\begin{array}{c}
E(\Gamma, \Gamma_0) = \sum_{bonds} K_r \left( r - r_0 \right)^2  + 
    \sum_{angles} K_{\theta} \left( \theta - \theta_0 \right)^2  + \\
    \sum_{dihedral} K_{\phi}^{(n)}  \left[1 + \cos \left( n \times (\phi - \phi_0)
    \right) \right] + \\
    \sum_{i < j -3} \{ \epsilon(i,j) [5 \left( \frac{\sigma_{ij}}{r_{ij}}
    \right)^{12}
    - 6 \left( \frac{\sigma_{ij}}{r_{ij}}\right)^{10} ] + \epsilon_2(i,j) \left(
    \frac{\sigma_{ij}}{r_{ij}}\right)^{12} \}.
\label{go_ham}
\end{array}
\ee

In the previous expression $r$ and $r_0$ represent the distances
between two subsequent residues at, respectively, the configuration
$\Gamma$ and the native state $\Gamma_0$. Analogously, $\theta$ 
($\theta_0$) and $\phi$ ($\phi_0$) represent the angles formed by
three subsequent residues and the dihedral angle defined by four
subsequent residues along the chain at the configuration $\Gamma$
($\Gamma_0$).  The dihedral potential consists of a sum of two terms
for every four adjacent $C_{\alpha}$ atoms, one with period $n =1$ and
one with $n=3$.  The last term in Eq. (\ref{go_ham}) contains the
non--local native interactions and a short range repulsive term for
non--native pairs (i.e. $\epsilon(i,j) = constant > 0$ and
$\epsilon_2(i,j)=0$ if $i$--$j$ is a native pair while $\epsilon(i,j) =
0$ and $\epsilon_2(i,j)= constant > 0$ if $i$--$j$ is a non--native
pair). The parameter $\sigma_{ij}$ is taken equal to $i$--$j$ distance
at the native state for native interactions, while $\sigma_{ij}=4$ \AA \
for non--native (i.e. repulsive) interactions.  Parameters $K_r$,
$K_{\theta}$, $K_{\phi}$, $\epsilon$ weight the relative strength of
each kind of interaction entering in the energy and they are taken to
be $K_r = 100 \epsilon$, $K_{\theta} = 20 \epsilon$, $K_{\phi}^{(1)} =
\epsilon$ and $K_{\phi}^{(3)} = 0.5 \epsilon$.  
With this choice of the parameters we found that the stabilizing energy 
residing in the tertiary contacts is approximately twice the stabilizing
energy residing in the torsional degrees of freedom. This balance among 
the energy terms is optimal to study the folding of our G\=o--like protein 
models.
The native contact map of a protein is derived with the CSU software based
upon the approach developed in ref. \cite{Sobolev96}. Native contacts
between pairs of residues $(i,j)$ with $j \leq i+3$ are discarded from
the native map as any three and four subsequent residues are already
interacting in the angle and dihedral terms.  A contact between two
residues $(i,j)$ is considered formed if the distance between the
$C_{\alpha}$'s is shorter than $\gamma$ times their native distance
$\sigma_{ij}$. It has been shown \cite{Onuchic99} that the results are
not strongly dependent on the choice made for the cut--off distance
$\gamma$.  In this work we used $\gamma = 1.2$.
We have used Molecular Dynamics (entailing the numerical integration of Newton's
laws of motion) for simulating the kinetics of the protein models.
We employed the simulation package AMBER (Version 4.1) \cite{Amber41:95} at 
constant temperature, i.e. using Berendsen algorithm for coupling the system to 
an external bath \cite{Berendsen84}.
Both temperature and energy are measured in units of the folding temperature 
$T_f$ in the simulations.  

For each protein model, several constant temperature simulations were 
made and combined using the WHAM algorithm 
\cite{Ferrenberg88,Ferrenberg89,Swendsen93} to generate a
specific heat profile versus temperature and a free energy $F(Q)$
as a function of the folding reaction coordinate Q. 
This algorithm is based on the fact that the logarithm of probability 
distribution $P(Q)$ of the values taken by a certain variable Q (e.g. the order 
parameter) at fixed temperature T may serve as an estimate for the the free 
energy profile $F(Q)$ at that temperature.
In fact, the probability to have a certain value $Q_1$ for the variable Q, at
temperature $T = 1/\beta$, in the canonical ensemble is given by:
\begin{equation}
P_{\beta}(Q_1) = \frac{W(Q_1) e^{-\beta E(Q_1)}}{Z_{\beta}}
\end{equation}
where $W(Q)$ is the density of configurations at a point $Q$ in the 
configurational space,
$Z_{\beta}$ is the canonical partition function at temperature $T=1/\beta$ and 
$E(Q)$ is the energy of the system at the value Q of the reaction 
coordinate\footnote{Since our model is almost energetically unfrustrated, the
energy fluctuations for a set of configurations with fixed Q are strongly reduced 
such that the energy in a given configuration could be considered as a function of $Q$.}.
Since the free energy $F$ is 
\begin{equation}
F(Q) = E(Q) - T S(Q)
\end{equation}
and the entropy S(Q) is related to the configurational density $W(Q)$
\begin{equation}
W(Q) \sim e^{S(Q)/k}
\end{equation}
where $k$ is the Boltzmann constant, it follows that
\begin{equation}
\frac{P_{\beta}(Q_1)}{P_{\beta}(Q_2)} = \frac{e^{-\beta F(Q_1)}}{e^{-\beta F(Q_2)}}
\end{equation}
and free energy differences can be computed by
\begin{equation}
-\beta (F(Q_1)- F(Q_2)) = \log{\frac{P_{\beta}(Q_1)}{P_{\beta}(Q_2)}}.
\end{equation}
By using the procedure of refs. \cite{Ferrenberg88,Ferrenberg89,Swendsen93}, 
data from a finite set of simulations can be used to obtain complete thermodynamic 
information over a large parameter region.

Probability distributions are obtained by sampling the configurational space
during Molecular Dynamics simulations.

For the smaller proteins ({\it CI2} and {\it SH3}) we have determined the errors 
on the estimates of the transition temperature and 
contact probabilities (or $\Phi$ values). This has been accomplished
by computing these quantities from several (more than 10) uncorrelated 
sets of simulations. We found that the standard deviation 
for each single contact probability is 0.06 for CI2 and 0.05 for SH3, while the 
transition temperature is determined in both cases with an uncertainty 
smaller than 0.5\%. These errors are obtained using about 200 uncorrelated conformations
in the transition state ensemble.
Since {\it Barnase}, {\it RNase H} and {\it CheY} have twice to three times the number
of tertiary contacts of {\it SH3} and {\it CI2}, in order
to have appropriate statistics, 
we have sampled about 500 uncorrelated conformations (thermally weighted)
for every transition state ensemble or intermediate.

\end{appendix}
\newpage

\subsection*{ Captions to the figures}

{\bf Fig. 1.}
(a) Free energy $F(Q)$ as a function of the reaction coordinate $Q$
around the folding temperature for the model of {\it CI2}. Free energies are
measured in units of $k_{B}T_f$, where $T_f$ is the folding temperature.
The unfolded, folded and transition 
state regions are shown in the light blue shaded areas.
(b) 
A typical sample simulation at a temperature around the folding temperature. 
The reaction coordinate Q as a function of time (measured in arbitrary 
unit of molecular dynamics steps) is shown. 
The two--state behaviour is apparent from the data.
The unfolded and folded 
states are equally populated at the folding temperature. 
(c) Heat capacity as a function of the temperature (units of folding
temperature).

{\bf Fig. 2.}
The results for the transition state structure from the simulations for CI2.
The probability of native contact formation at the transition state (left panel), and
bond $\Phi$--values (right panel) are shown. Different colors
indicate different values from 0 to 1, as quantified by the color
scale. The $\alpha$--helix, the interactions between the strands 4 and 5, and the
minicore (i.e. interactions between residues 32,38 and 50) are the
parts of the structure formed with the highest probability, although they
are not fully formed. Overall, the  transition state ensemble
appears as an expanded version of the native state where most contacts
have a similar probability of participation, but some interactions are
less like to occur. These results agree with the transition state
structure experimentally obtained.

\begin{figure}[h]
\centerline{\psfig{figure=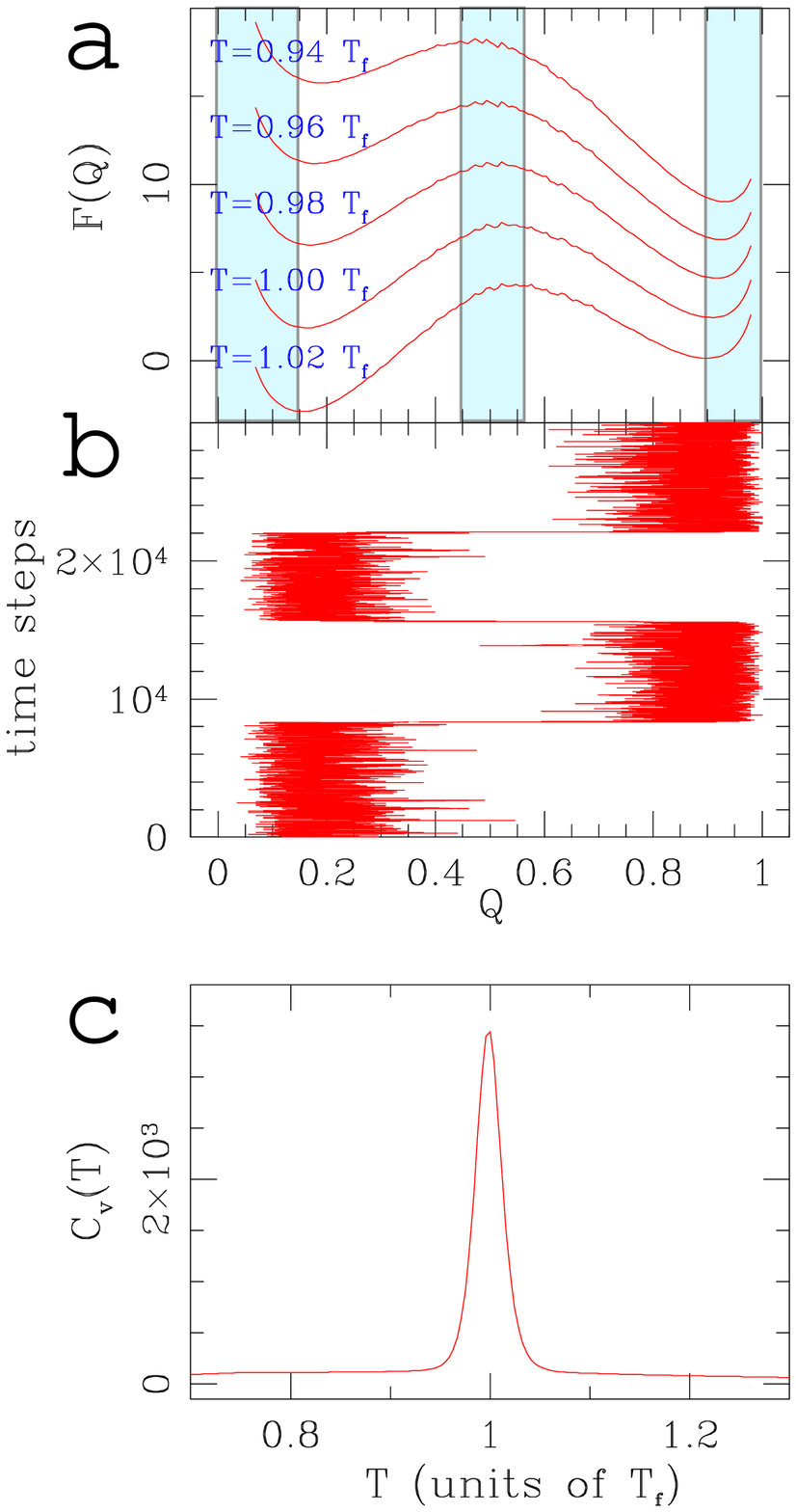,height=19.0cm,clip=}}
\caption{
}
\label{fig1_ci2}
\end{figure}

\begin{figure}[h]
\centerline{
\psfig{figure=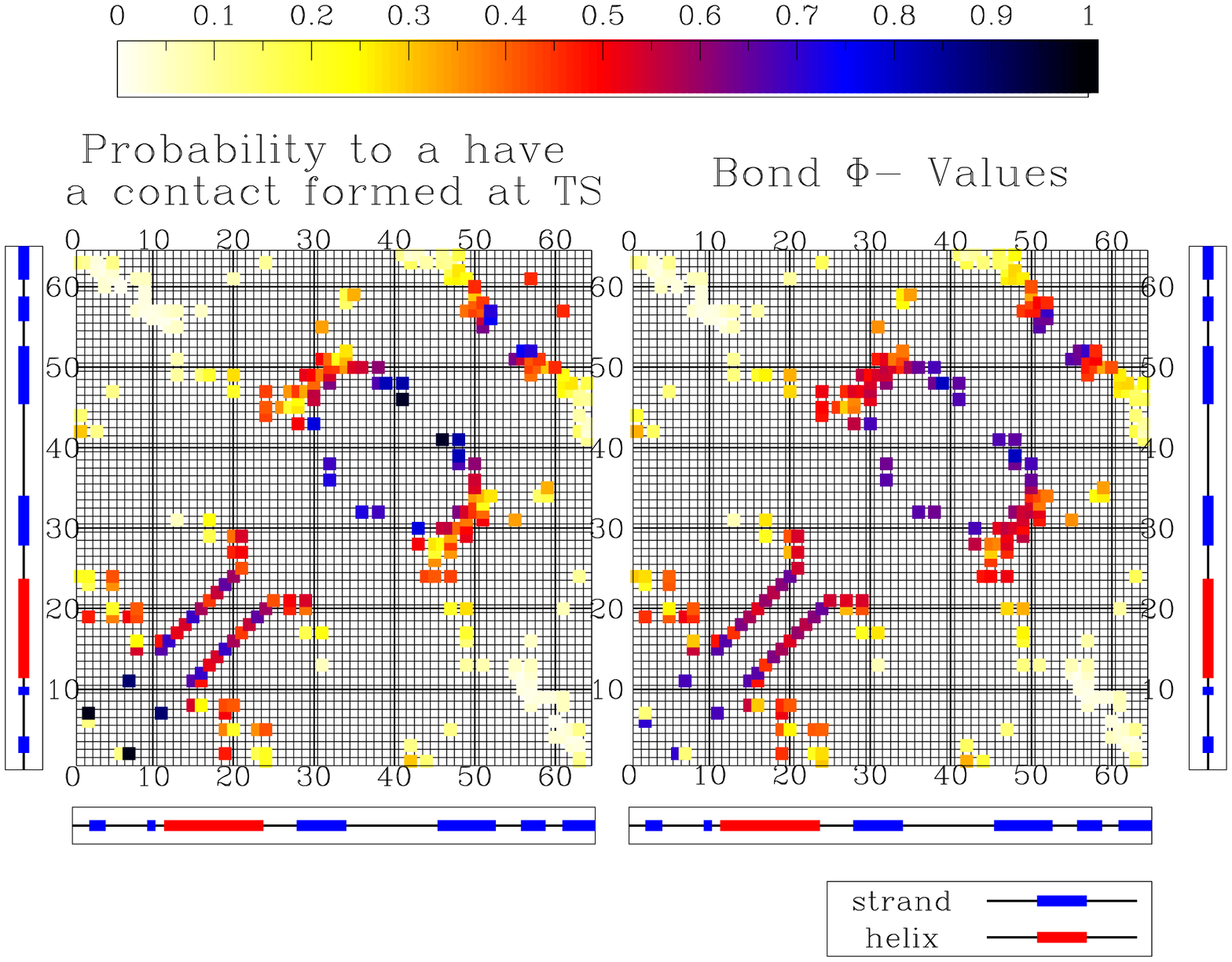,height=15cm,clip=}}
\caption{ 
}
\label{fig2_ci2}
\end{figure}

\begin{figure}[h]
\centerline{\psfig{figure=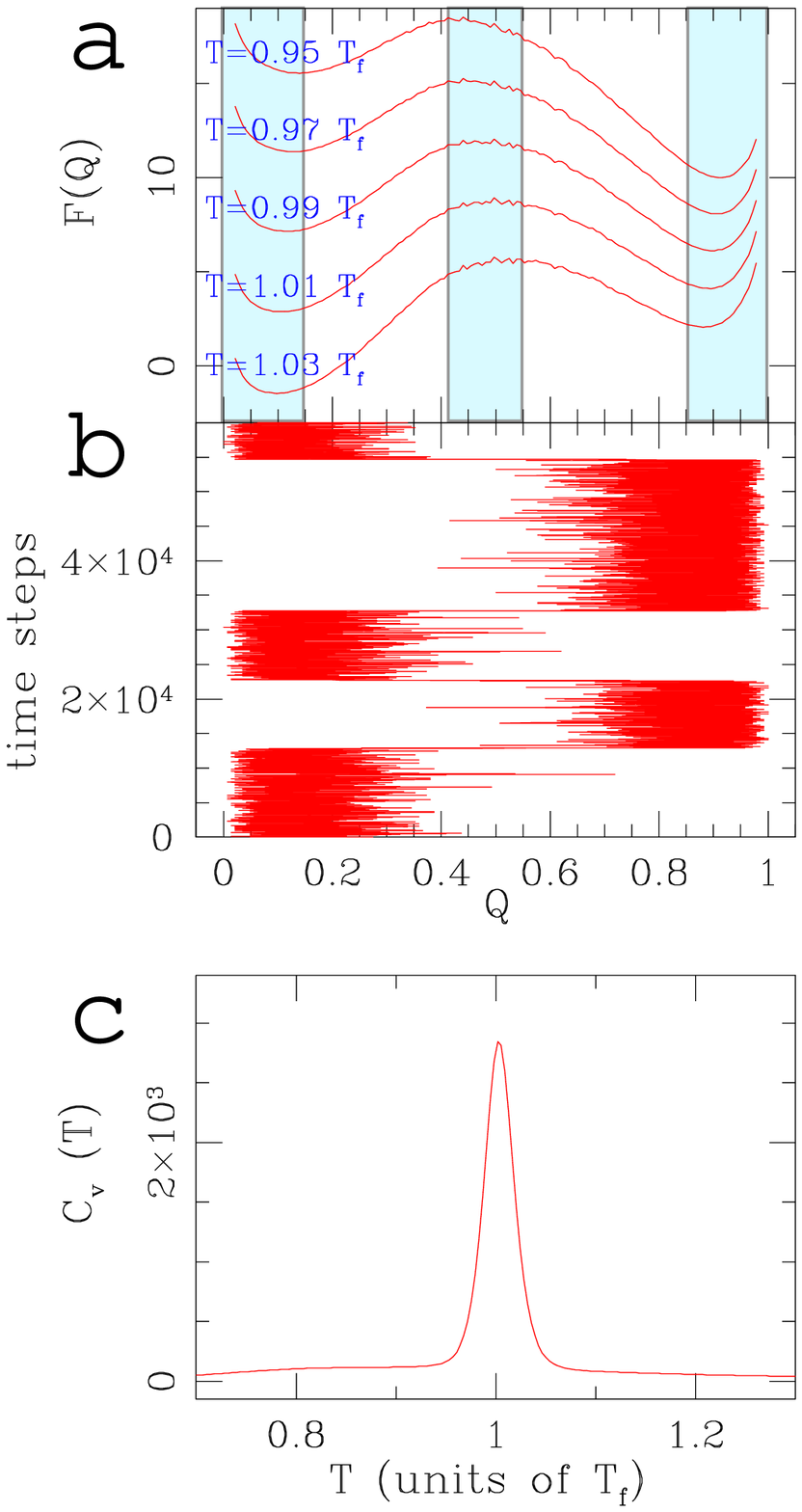,height=19.0cm,clip=}}
\caption{
}
\label{fig1_sh3}
\end{figure}

\begin{figure}[h]
\centerline{
\psfig{figure=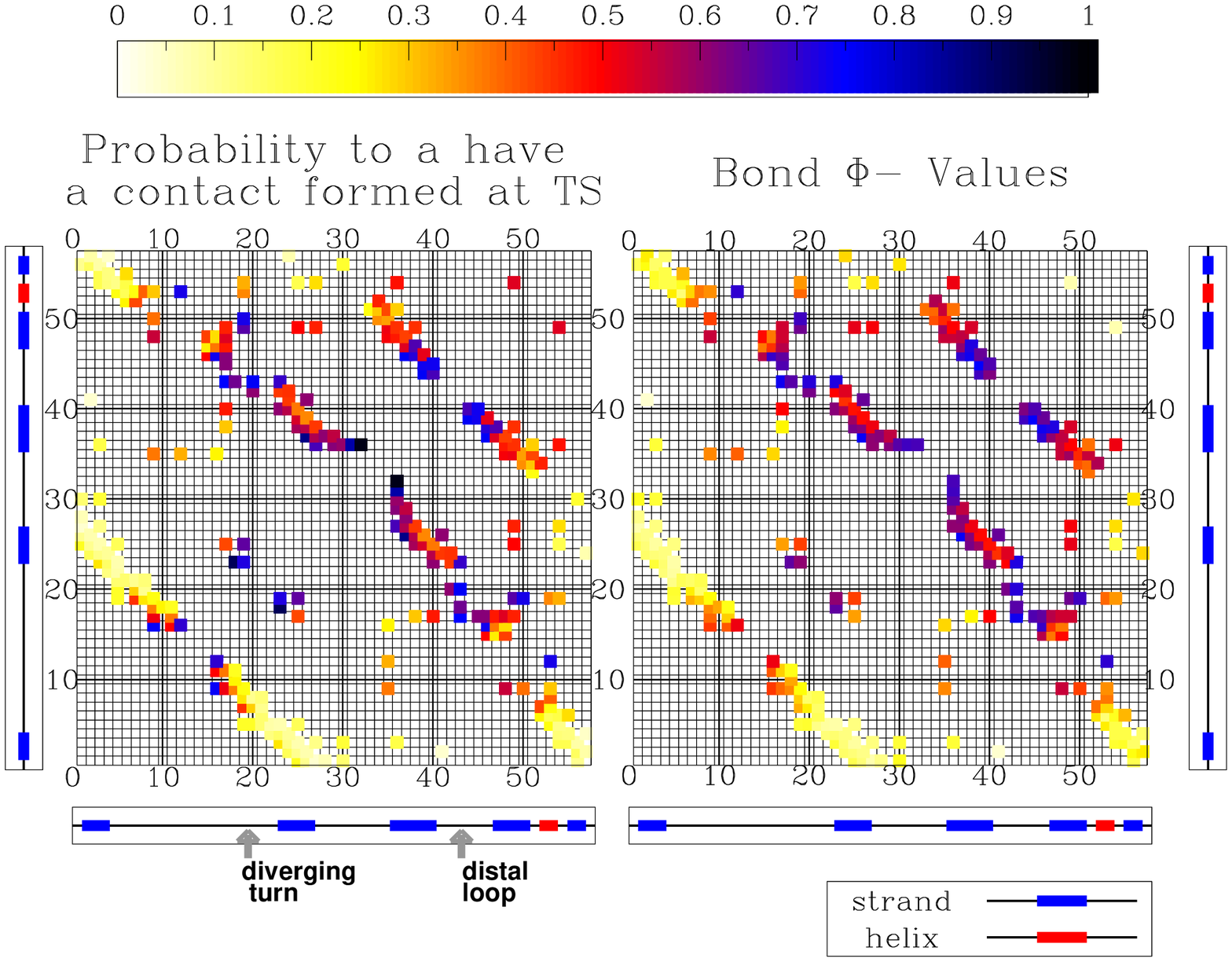,height=15.0cm,clip=}}
\caption{ 
}
\label{fig2_sh3}
\end{figure}

{\bf Fig. 3.}
(a) Free energy $F(Q)$ as a function of the reaction coordinate $Q$ for a
set of temperatures around the folding temperature. Free energies are
measured in units of $k_{B}T_f$.  The choices for the
unfolded, folded and transition state regions are marked as shaded regions.
(b) The reaction coordinate Q as a function of time (unit of molecular dynamics 
steps), from a typical sample simulation around the folding temperature. As
in Fig. \protect\ref{fig1_ci2}, the two--state behaviour is apparent. 
At the
transition temperature the model protein has equal probability to be found in the
unfolded or in the folded state.
(c) Heat capacity as a function of the temperature, in units of folding temperature.

{\bf Fig. 4.}
The transition state structure as obtained from the simulations for SH3.
Panel in the left represents the probability for a native contact to be
formed at the transition state, while the panel in the right shows the results for
bond $\Phi$--values. Different colors indicate different values from 0
to 1, as quantified by the color scale. Diverging turn and distal loop 
are marked on the contact map.  The interactions within and
between these two parts of the protein chain appear to be formed with high
probability. The interactions between the two strands joined by the
distal loop are partially formed, while the contacts involving the
first 20 residues do not contribute to the transition state structure.  This
description of the transition state is in agreement with experimental
results.

{\bf Fig. 5.}
(a) Free energy $F(Q)$ of {\it Barnase} protein as a function of the reaction 
coordinate $Q$ around the folding temperature. 
Free energies are measured in units of $k_{B}T_f$.
The unfolded, folded and 
intermediate  state regions are marked in green, while the top of the two barriers 
are marked in light blue. The local minimum in the free energy profile
between the unfolded and folded minima locates the folding intermediate 
state. The presence of a folding intermediate state is also evident from panel (b),
where the order parameter $Q$ is plotted as a function of time for a typical molecular 
dynamics simulation around the folding temperature. In the interval $Q \in (0.4 - 0.5)$,
the same state (i.e. with the same average structure) is visited both from the 
unfolded and folded structures.

{\bf Fig. 6.}
The probability of native contact formation for the intermediate (left panel) and 
transition state (right panel) structures as obtained from our simulations of Barnase.
Different colors indicate different values from 0 to 1, as quantified by 
the color scale. 
The earliest formed part of the protein appears to be the $\beta$--sheet region,
in agreement with experimental results. The core 3 (formed by  
loops 3 and 5 to the $\beta$--sheet) is formed at the intermediate and 
transition state, while core 1 (the packing of the helix 1 against 
the $\beta$--sheet) and the core 2 (the interactions between the
hydrophobic residues from the helices 2 and 3, the strand 1,
and the first two loops) start to form only after the transition state. 
The formation of the $\alpha$--helix 1 occurs as
a late event of the folding from our simulations, while from experimental 
results it seems to be already formed at the intermediate and transition state. The 
early formation of the 
$\alpha$--helix is most probably due to energetic factors rather then 
from topology requirements (and then beyond the prediction possibility 
of this model), as detailed in the text.

{\bf Fig. 7.}
(a) Free energy $F(Q)$ of the model of {\it RNase H} as a function of the reaction 
coordinate $Q$ around the folding temperature. 
Free energies are measured in units of $k_{B}T_f$.
The regions corresponding to the
unfolded, folded and intermediate  state are marked in green, while the top of 
the two barriers are marked in light blue. A folding intermediate is detected as a
local minimum in the free energy between the unfolded and folded minima.
In panel (b) the fraction of native contacts formed, $Q$, is plotted versus the 
simulation time for a sample of our simulations (at a temperature $T=0.99 T_f$) 
where the transition from unfolded 
to folded state is observed. The local minimum of panel (a) corresponds to a 
transiently populated intermediate (located at $Q$ around 0.4) that later evolves to the fully 
folded state.

{\bf Fig. 8.}
The probability of native contact formation at the intermediate (left panel) and 
transition state (right panel) structure, as observed for the {\it RNase H} model. 
Different colors indicate different values from 0 to 1, as quantified by 
the color scale. 
In agreement with experimental results, we found that 
interactions  involving the $\alpha$--helix 1 are the first formed in the folding process.
Contacts between the $\alpha$--helix 1 and the strand 4 are highly probably formed at the
intermediate. Also the $\alpha$--helix 4 is well structured and the $\beta$--sheet is 
partly formed.
These interactions strengthen at the transition state where also the $\beta$--sheet is 
almost completely formed, while the packing of helix 5 across the sheet is not yet accomplished.

\begin{figure}[h]
\centerline{\psfig{figure=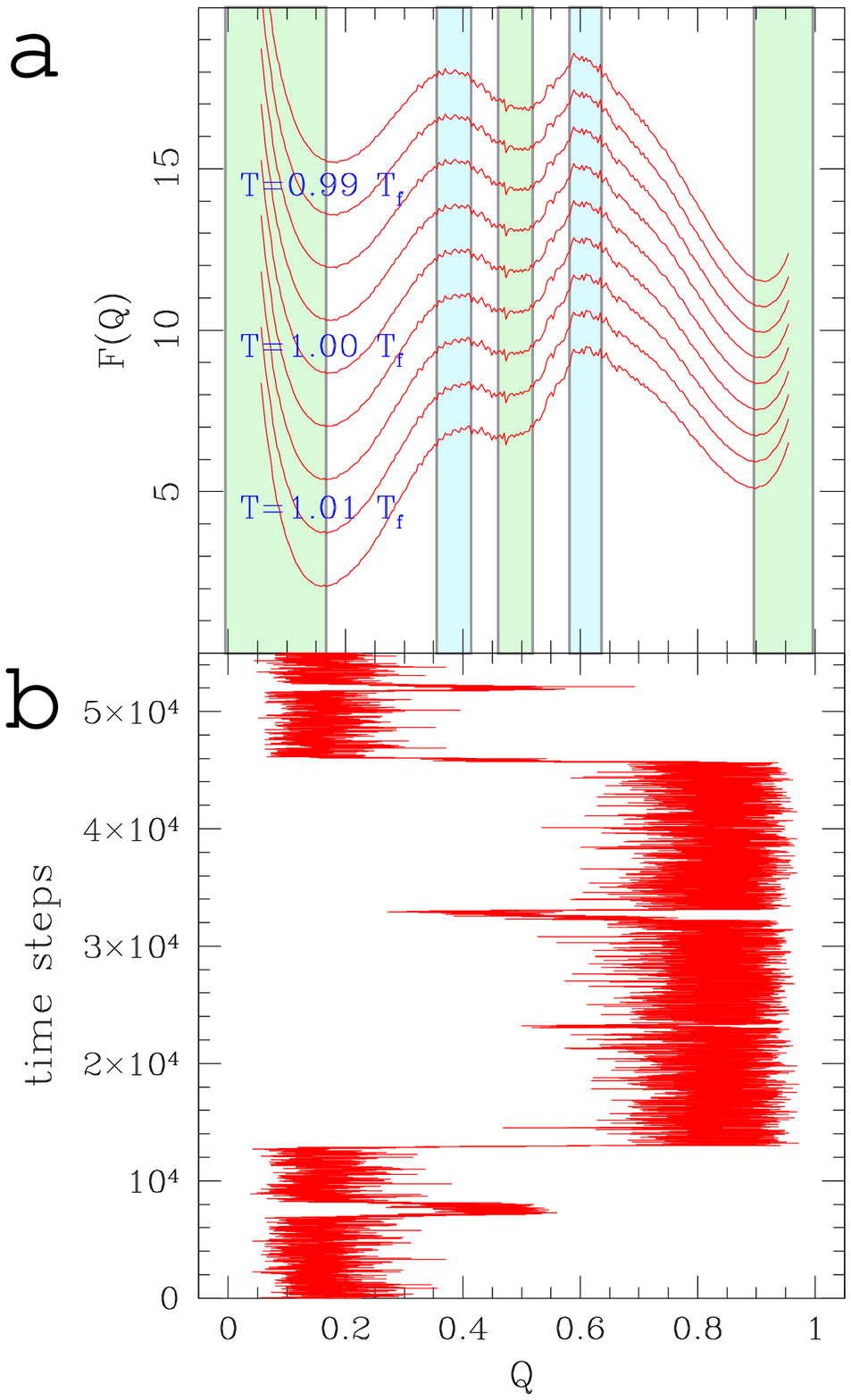,height=19.0cm,clip=}}
\caption{
}
\label{fig1_brn}
\end{figure}

\begin{figure}[h]
\centerline{
\psfig{figure=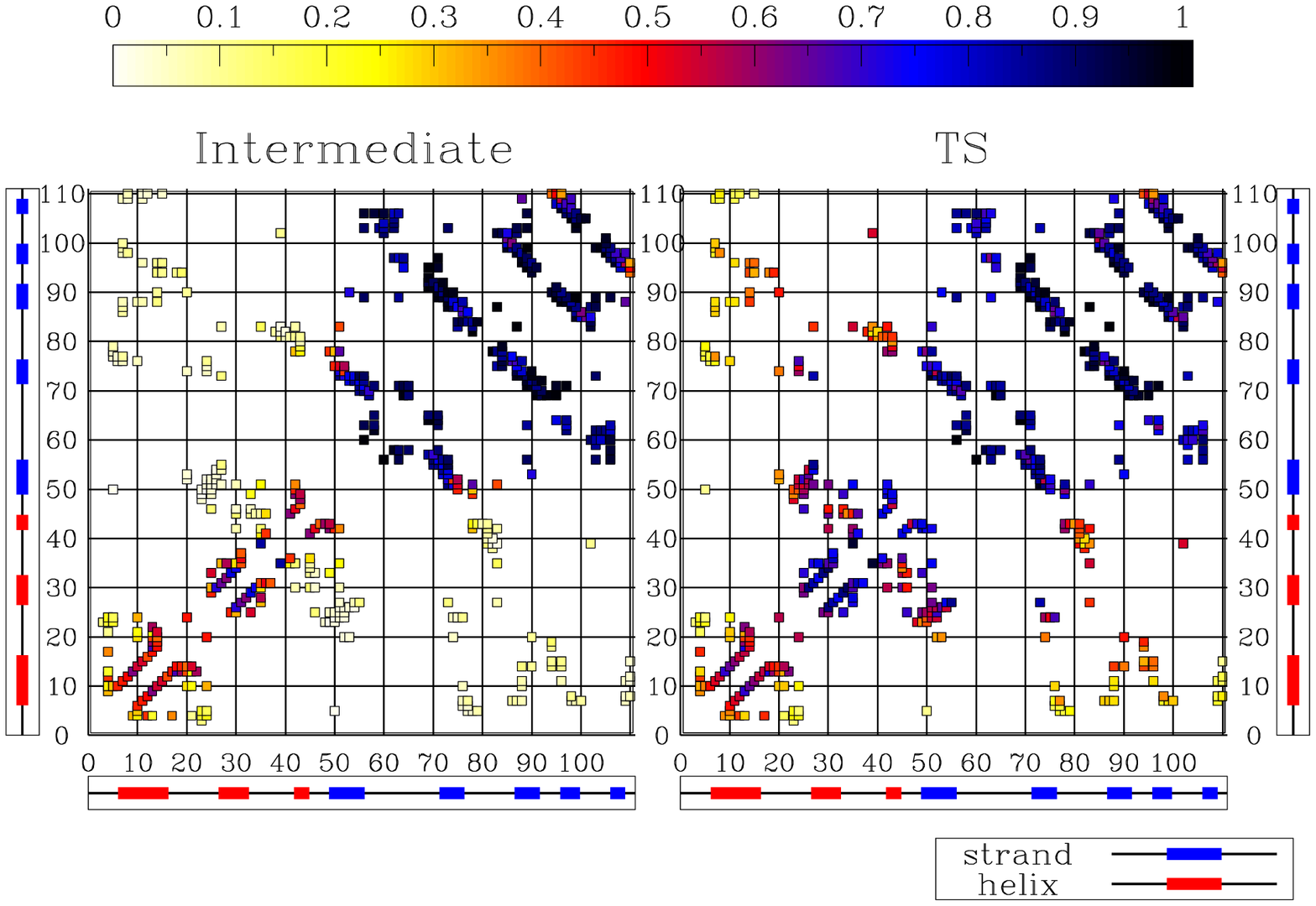,height=15.0cm,clip=}}
\caption{ 
}
\label{cm_stat.brn.fig}
\end{figure}

\begin{figure}[h]
\centerline{\psfig{figure=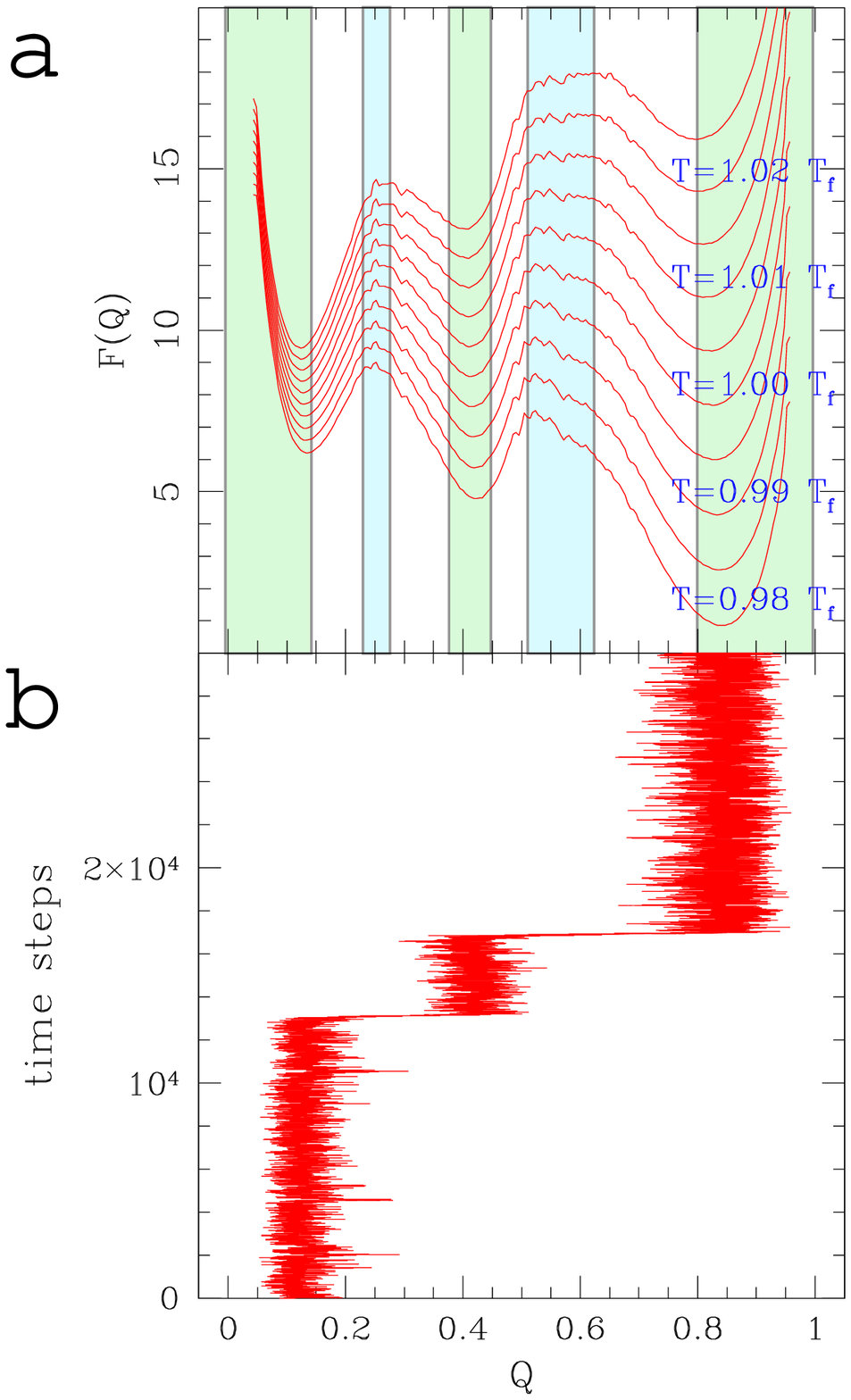,height=19.0cm,clip=}}
\caption{
}
\label{fig1_rnase}
\end{figure}

\begin{figure}[h]
\centerline{\psfig{figure=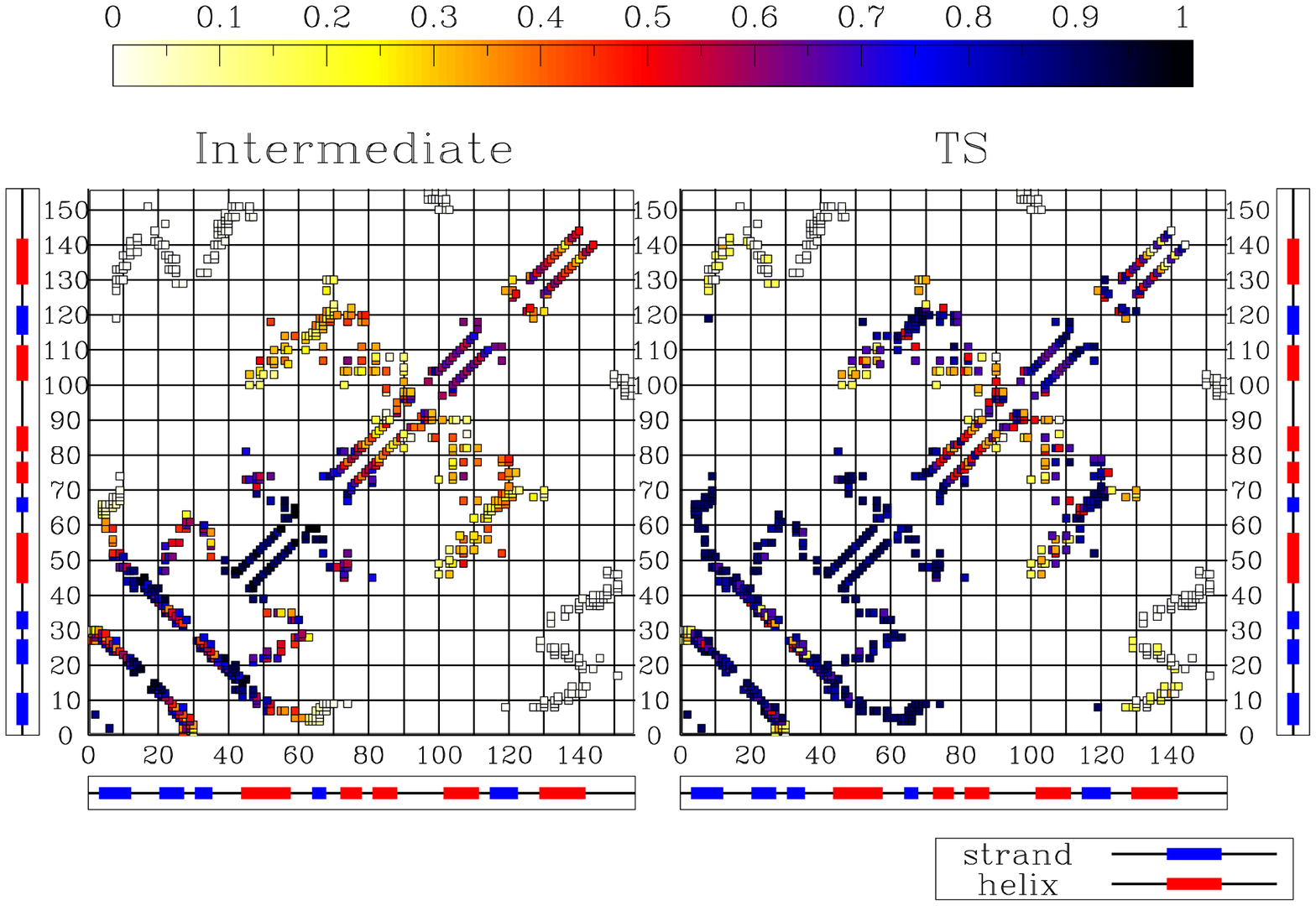,height=15.0cm,clip=}}
\caption{ 
}
\label{fig_rnase_cmap}
\end{figure}

\begin{figure}[h]
\centerline{\psfig{figure=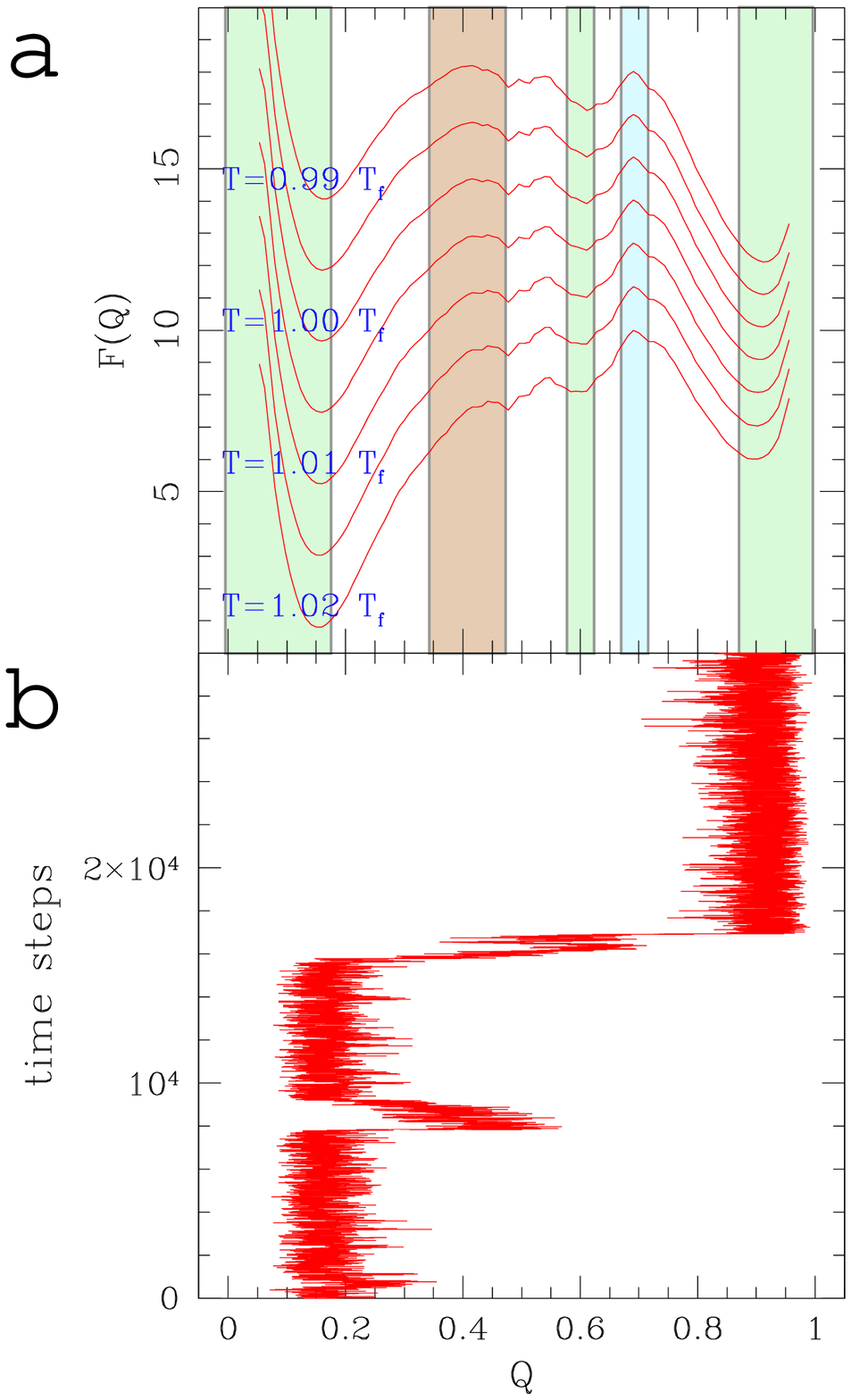,height=19.0cm,clip=}}
\caption{
}
\label{fig1_chey}
\end{figure}

\begin{figure}[h]
\centerline{\psfig{figure=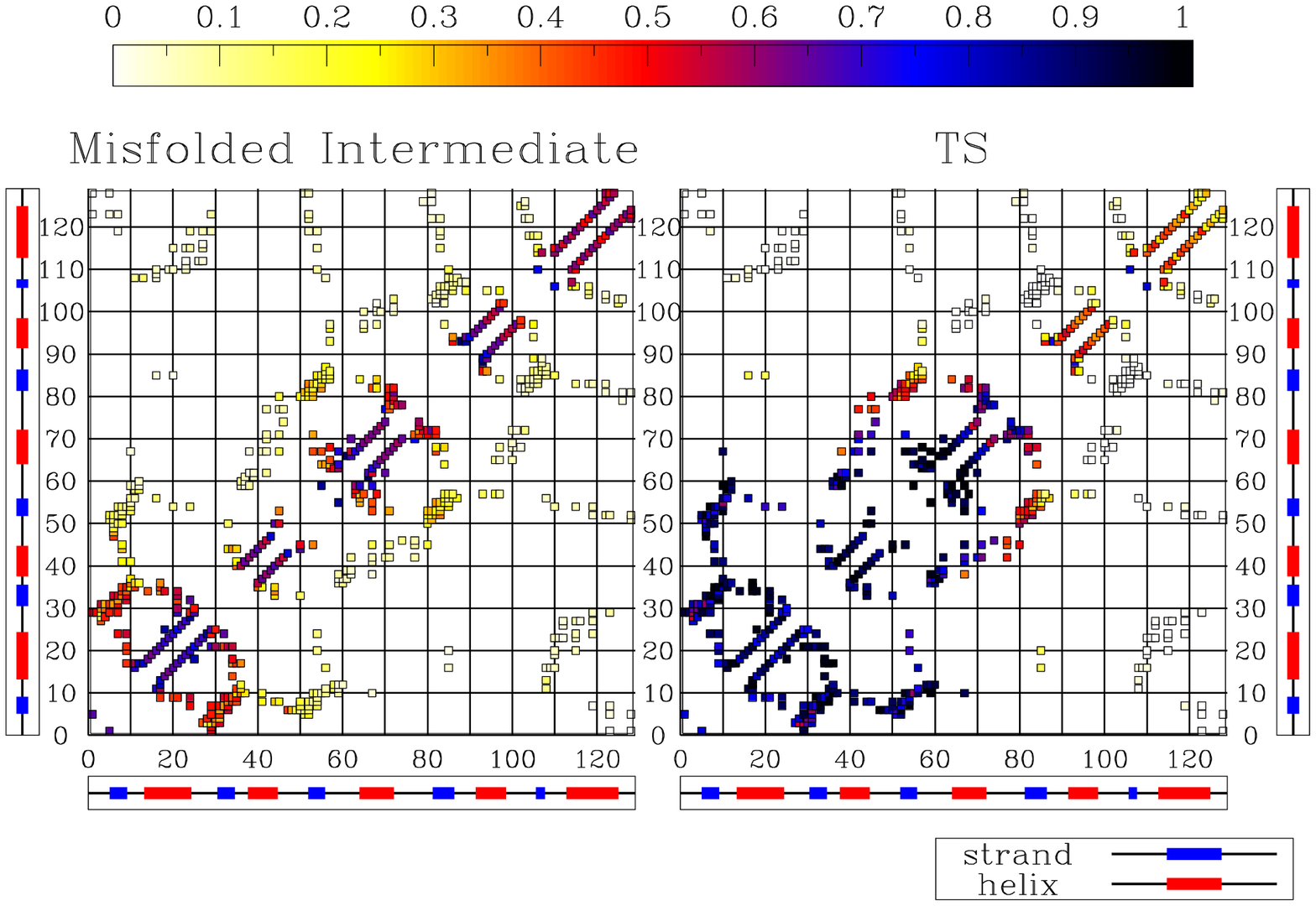,height=15.0cm,clip=}}
\caption{ 
}
\label{fig_chey_cmap}
\end{figure}

{\bf Fig. 9.}
(a) Free energy $F(Q)$ profile for the model of {\it CheY} plotted as a function of 
the reaction coordinate $Q$ for a set of temperatures around the folding temperature. 
Free energies are measured in units of $k_{B}T_f$.
Differently from the corresponding figures of {\it Barnase} (Fig. \protect\ref{fig1_brn}) and
{\it RNase H} (Fig. \protect\ref{fig1_rnase}), two different structures are populated between 
the folded and unfolded states. In addition to the ``on--route" intermediate state 
(marked in green as the regions corresponding to the folded and unfolded states),
a ``misfolded" intermediate structure (marked in brown at Q around 0.4)
is transiently visited from the unfolded state. 
The top of the two barriers are marked in light blue.
In agreement with experimental results, we found
that in this ``misfolded" structure, all the five  $\alpha$--helices are 
rather structured while, in the later occurring ``on--route" intermediate and 
transition state ensemble, the helices 4--5 are completely  unstructured 
(see fig. \protect\ref{fig_chey_cmap}).
Panel (b) shows a typical sample of the simulation around the folding temperature, 
in a region where the folding occurs. The first transiently populated intermediate state
corresponds to a structure where all the helices are formed. Before to proceed
to the folded state, a partial unfolding occurs.

{\bf Fig. 10.}
The probability  of the native {\it CheY} contacts to be formed in the ``misfolded"
intermediate (left panel) and transition state (right panel)
for the model protein.
Different colors indicate different values from 0 to 1, as quantified by 
the color scale. 
In agreement with experimental data, all the helices are mostly formed 
in the transiently populated ``misfolded" structure, while helices 4 and 5 
are rather unstructured at the transition state.
The two subdomains experimentally detected in the {\it CheY} transition state 
\protect\cite{LopezHernandez96,LopezJMB97}
are evident in the figure:
the first part of the protein (all interactions arising from the $\alpha$--helices 1--2 and 
the $\beta$--strands 1--3) 
is folded, while the second part (interactions among the $\alpha$--helices 4--5 and
the $\beta$--strands 4--5) is completely unfolded. The helix 3 is structured but the interactions
between the helix 3 and the rest of the protein are not completely formed.

\newpage


\newpage
\onecolumn

\pagestyle{empty}
\thispagestyle{empty}

\end{document}